\documentclass[pdflatex,sn-mathphys-num]{sn-jnl}

\usepackage{graphicx}%
\usepackage{multirow}%
\usepackage{amsmath,amssymb,amsfonts}%
\usepackage{amsthm}%
\usepackage{mathrsfs}%
\usepackage[title]{appendix}%
\usepackage{xcolor}%
\usepackage{textcomp}%
\usepackage{manyfoot}%
\usepackage{booktabs}%
\usepackage{algorithm}%
\usepackage{algorithmicx}%
\usepackage{algpseudocode}%
\usepackage{listings}%

\theoremstyle{thmstyleone}%
\theoremstyle{thmstyletwo}%

\theoremstyle{thmstylethree}%

\begin{document}

\title[Article Title]{Apparatus for the measurement of birefringence maps of optical materials: the case of crystalline silicon for Einstein Telescope}


\author[1,2]{\fnm{Alina Mariana} \sur{Soflau}}
\equalcont{now at: NIKHEF, Science Park 105, Amsterdam, 1098 XG, NL.}
\author*[3,4]{\fnm{Federico} \sur{Della Valle}}\email{federico.dellavalle@unisi.it}
\author[1]{\fnm{Francesco} \sur{Cescato}}
\author[1,2]{\fnm{Giovanni} \sur{Di Domenico}}
\author[1,2]{\fnm{Aur\'elie} \sur{Mailliet}}
\author[1,2]{\fnm{Lorenzo} \sur{Malagutti}}
\author[3,4]{\fnm{Emilio} \sur{Mariotti}}
\author[1,2]{\fnm{Andrea} \sur{Mazzolari}}
\author[1,2]{\fnm{Marco} \sur{Romagnoni}}
\author[1,2]{\fnm{Guido} \sur{Zavattini}}

\affil[1]{\orgdiv{Dip. di Fisica e Scienze della Terra}, \orgname{University of Ferrara}, \orgaddress{\street{via Saragat 1, Edificio C}, \city{Ferrara}, \postcode{I-44122}, \country{Italy}}}
\affil[2]{\orgdiv{Sez. di Ferrara}, \orgname{INFN}, \orgaddress{\street{via Saragat 1, Edificio C}, \city{Ferrara}, \postcode{I-44122}, \country{Italy}}}
\affil[3]{\orgdiv{Dip. di Scienze Fisiche, della Terra e dell'Ambiente}, \orgname{University of Siena}, \orgaddress{\street{via Roma 56}, \city{Siena}, \postcode{I-53100}, \country{Italy}}}
\affil[4]{\orgdiv{Sez. di Pisa}, \orgname{INFN}, \orgaddress{\street{Largo B. Pontecorvo 3}, \city{Pisa}, \postcode{I-56127}, \country{Italy}}}


\abstract{Einstein Telescope (ET) is expected to achieve sensitivity improvements exceeding an order of magnitude compared to current gravitational-wave detectors. The rigorous characterization in optical birefringence of materials and coatings has become a critical task for next-generation detectors, especially since this birefringence is generally spatially non-uniform. A highly sensitive optical polarimeter has been developed at the Department of Physics and Earth Sciences of the University of Ferrara and INFN - Ferrara Section, Italy, aimed at performing two-dimensional birefringence mapping of substrates. In this paper we describe the design and working principle of the system and present results for crystalline silicon, a candidate material for substrates in the low-frequency (LF) interferometers of ET. We find that the birefringence is of order $10^{-7}$ for commercially available samples and is position dependent in the silicon (100)-oriented samples, with variations in both magnitude and axis orientation. We also measure the intrinsic birefringence of the (110) surface. Implications for the performance of gravitational-wave interferometers are discussed.}




\maketitle

\section{Introduction}\label{sec1}

Substrates and coatings exhibit birefringence. This optical property of materials is crucial in high-precision optical instruments sensitive to polarization effects, such as interferometers. When linearly polarised light crosses a birefringent medium, its polarization, in the most general case, becomes elliptical. In interferometric systems, this results in a polarization component orthogonal to the original one, which causes power losses due to imperfect constructive interference \cite{Winkler1994}. If non-uniform, birefringence can also distort the wavefront, generating higher-order modes that may further compromise the sensitivity of the system. Although existing room-temperature gravitational-wave interferometers employing fused silica substrates have not faced significant challenges related to birefringence, the cryogenic detector KAGRA \cite{KAGRA_2020} has demonstrated the potential impact of this effect. Specifically, sapphire used in KAGRA's test masses was found to exhibit higher than expected birefringence \cite{Hirose2020,Zeidler2023}, causing a significant power loss between s- and p-polarizations (6.1\% in the X arm and 11\% in the Y arm) when the arms were off-resonance \cite{KAGRA_2020}. Birefringence was also found to vary across the substrate, leading to changes in both the phase and amplitude of p-polarised light \cite{KAGRA_2020}. Third-generation gravitational-wave interferometers strive for higher sensitivity compared to LIGO, VIRGO and KAGRA; the study of static birefringence becomes therefore even more critical, as its full impact is still not completely understood.

Any birefringence denounces an anisotropy. This can have an intrinsic origin encoded in the structure of the material, and in this case can have a global character as for the case of quartz and sapphire, or can be associated with local defects like dislocations or residual internal stress. Birefringence can also be induced by any anisotropic field applied to the sample, like a force or a temperature gradient. A cubic crystal structure should guarantee complete optical isotropy, but birefringence and possibly dichroism also result from spatial dispersion and quadrupole transitions for light traveling in a cubic crystal along the [110] direction \cite{Lorentz1921,Agranovich1984}. This amounts to say that, at least in principle, a cubic structure has the optical properties of a uniaxial crystal. It has been shown that this small birefringence takes the value
\begin{equation}
\Delta n=0.44\,\pi\,\overline{n}\,(\overline{n}^2-1)^2\,\frac{a^2}{\lambda^2}
\label{Lorentz}
\end{equation}
where $\overline{n}$ is the average value of the index of refraction, $a$ is a length of the order of the lattice constant and $\lambda$ the wavelength of the light.
For the European third-generation gravitational-wave detector Einstein Telescope (ET), crystalline silicon, having a cubic face-centered structure, has been proposed as the substrate material due to its extremely low optical absorption for wavelengths $\lambda \gtrsim 1550$~nm and mechanical dissipation at low temperature. However, even disregarding spatial dispersion (with $a\approx1$~\AA~the above formula gives $\Delta n\approx5.8\times10^{-6}$ @ 1064~nm and $\Delta n\approx2.5\times10^{-6}$ @ 1550~nm) Einstein Telescope requires large-scale mirrors with diameter and thickness about 0.5~m that have never before been realised and tested before. Hence an extended measurement campaign could help manufacturers in the quest for growing more and more perfect crystals with the right orientation. The work presented here is a first step in this direction.

The experimental polarimetric approach presented in the paper follows the same principles as the VMB\-@\-CERN experiment \cite{Zavattini2022}, originally designed to measure the magnetic birefringence of vacuum. In Section~\ref{sec:Polarimetry} the physical principles of the method are detailed; in this context, a spurious rotation appears in addition to ellipticity. The experimental scheme is discussed in Section~\ref{sec:Experimental}, with particular attention to spurious ellipticity signals and to calibration. The experimental results are presented and discussed in Section~\ref{sec:Results}. Possible developments of the technique are discussed in the Conclusions section. 

\section{Principles of polarimetry}\label{sec:Polarimetry}

In the most general case, the index of refraction of a medium is a complex quantity: $\widetilde{n}= n + i \kappa$. The real part $n$ (commonly referred to as the index of refraction $\textit{tout court}$) is the factor by which the speed and the wavelength of the radiation are reduced with respect to their vacuum values. The imaginary part, $\kappa$, known as the extinction coefficient, describes the absorption of the medium. For two orthogonal directions, denoted as $\parallel$ and $\perp$, if the difference in refraction indices is non-zero, i.e. $\Delta n= n_{\parallel}-n_{\perp} \neq 0$, then the medium is said to manifest a linear birefringence. Similarly, a non-zero difference in the absorption indices $\kappa$ along the same axes implies the presence of a linear dichroism $\Delta\kappa=\kappa_{\parallel}-\kappa_{\perp}$. In a single pass through a linear birefringent medium, light acquires an ellipticity, whereas a linear dichroic medium produces a rotation of the polarization. As will be seen in the following, a rotation also results from interference in a birefringent \'etalon.

\subsection{Linear birefringence and dichroism: ellipticity and rotation}

\begin{figure} [bht]
  \centering 
  \includegraphics[width=8.0 cm]{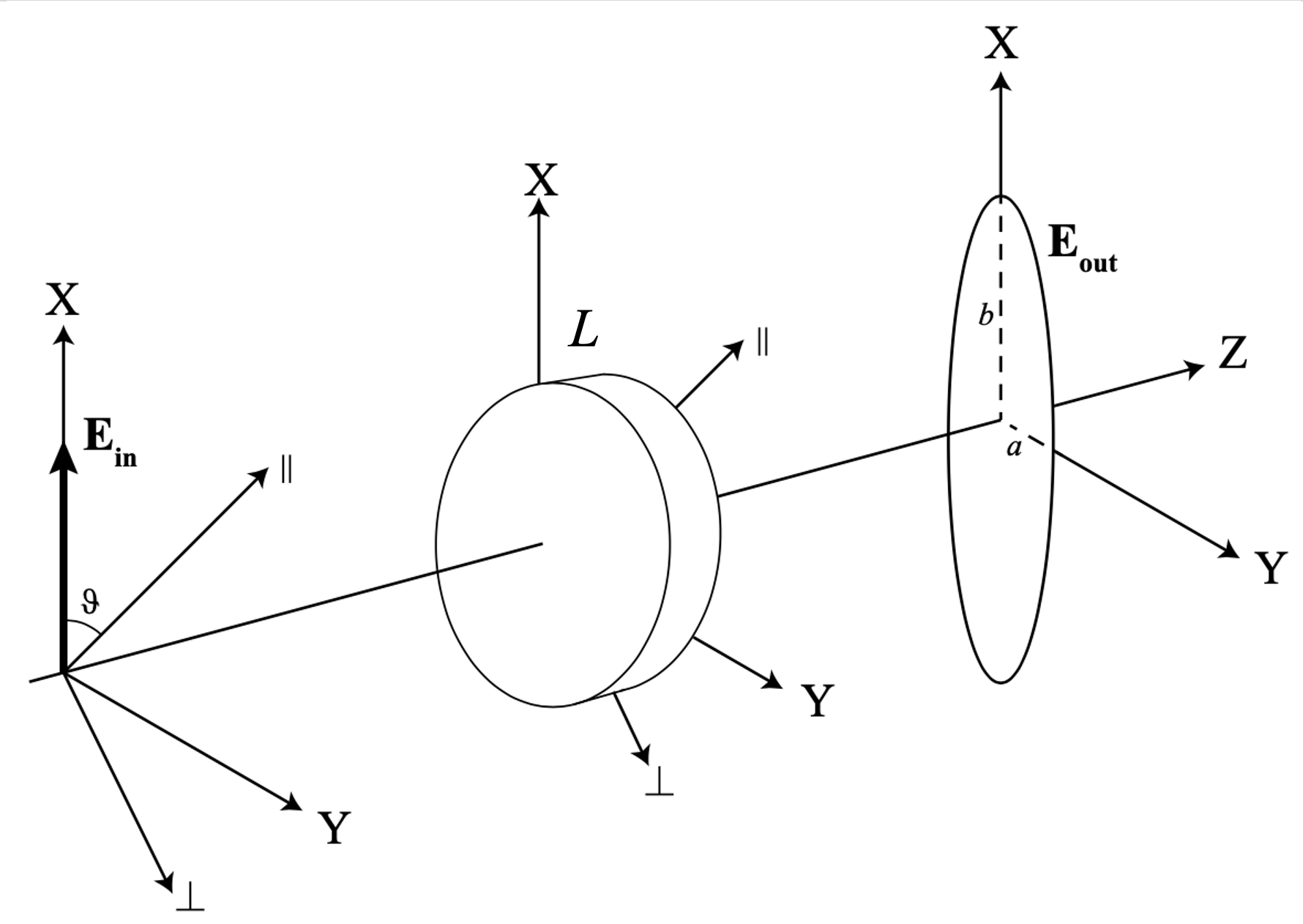}
  \caption{ Reference frame for the birefringence calculations. X and Y are the axes of the laboratory frame, whereas $\parallel$ and $\perp$ are the slow and fast (or extraordinary and ordinary) axes of the birefringent medium, respectively.}
  \label{fig:schemabirif}
\end{figure}

We begin considering a linearly birefringent medium and a linearly polarised light beam. With reference to Figure~\ref{fig:schemabirif}, suppose that the beam $\vec{E}_{\rm in}=E_{0}\,\hat{X}$ is linearly polarised along $X$ and propagates along the $Z$ axis through a uniformly birefringent medium with thickness $L$ and both slow ($\parallel$, extraordinary) and fast ($\perp$, ordinary) axes perpendicular to $Z$ ($n_\parallel>n_\perp$). Finally, let the slow axis of the medium form an angle $\vartheta$ with the $X$ axis. The electric field before entering the medium can be projected onto the axes $\parallel$ and $\perp$: 
\begin{equation}
 E_{\parallel}=E_0\cos\vartheta, \qquad\qquad E_{\perp}=-E_0\sin\vartheta .
\end{equation}

After crossing the birefringent medium, these components acquire a phase difference $\Delta \phi$, and the  two output field components become: 
\begin{equation}
    E'_{\parallel}= E_0\,e^{i\Delta\phi/2}\cos\vartheta,\qquad\qquad E'_{\perp}=-E_0\,e^{-i\Delta\phi/2} \sin\vartheta,
\end{equation}
where $\Delta\phi$ is defined in terms of the optical path difference $\Delta \mathcal{D}$ between the $\parallel$ and $\perp$ components of the electric field: 
   \begin{equation}
    \Delta\phi=\frac{2\pi}{\lambda}\int_L\Delta n(z) \,dz= \frac{2\pi}{\lambda} \Delta \mathcal{D}.
    \label{eq:phasedifference}
  \end{equation} 
  
The output field in the laboratory reference frame, assuming $\Delta\phi\ll 1$ (a condition that holds consistently throughout this work), is given by the matrix product: 
\begin{equation}
  \vec{E}_{\rm out}=
  \begin{pmatrix}E'_{X}\\E'_{Y}\end{pmatrix}=
       \begin{pmatrix}\cos\vartheta &-\sin\vartheta \\
    \sin\vartheta&\cos\vartheta
      \end{pmatrix}
      \begin{pmatrix}E'_\parallel\\E'_\perp
          \end{pmatrix}= E_0
      \begin{pmatrix}
          1+i\dfrac{\pi}{\lambda}\Delta \mathcal{D}\cos2 \vartheta \\
          i \dfrac{\pi}{\lambda} \Delta \mathcal{D} \sin 2 \vartheta
      \end{pmatrix}.
   \label{eq:outputfield}
    \end{equation}
    
The passage of light through a birefringent medium implies then that light acquires an electric field component $E'_{Y}$  phase-shifted by $\pi/2$ and orthogonal to the incident field. To first order, $\vec{E}_{\rm out}$ describes an ellipse. The ratio of the minor to the major axis of the ellipse is the {\it ellipticity}
\begin{equation}
    \psi(\vartheta)\approx i \dfrac{\Delta \phi}{2}\,\sin 2 \vartheta =  i\dfrac{\pi} {\lambda} \int_L\Delta n(z)\,dz\,\sin 2 \vartheta=  i\dfrac{\pi} {\lambda} \Delta \mathcal{D}\,\sin 2\vartheta=\pm i\dfrac{a}{b}.
    \label{eq:elliptdeltan}
\end{equation}

Ellipticity can be treated as an imaginary quantity, with the sign distinguishing the two rotation directions. The Jones matrix representation of a  birefringent medium with slow and fast axes rotated by an angle $\vartheta$ relative to the laboratory frame is, for small $\Delta\phi$,
\begin{align}
 \textbf{BF}(\vartheta) \simeq
    \begin{pmatrix}
     \displaystyle 1+i\frac{\Delta\phi}{2}\cos2\vartheta & \displaystyle i\frac{\Delta\phi}{2}\sin2\vartheta \\\\
     \displaystyle i\frac{\Delta\phi}{2}\sin2\vartheta & \displaystyle 1-i\frac{\Delta\phi}{2}\cos2\vartheta 
     \label{eq:bf}
   \end{pmatrix}.
\end{align}

To determine a birefringence $\Delta n$, the electric field of light in a direction orthogonal to its initial polarisation must be measured. 
One must pay attention to the fact that ellipticity, being an integral quantity, allows only to measure an average value of the birefringence: several birefringent slabs each having a small phase difference and an arbitrary orientation combine together, and the system is equivalent to a single slab with well-defined phase difference $\Delta\phi$ and orientation $\vartheta$ \cite{Brandi1997APB}. It is also important to note that, for a uniformly birefringent medium, suppression of birefringence effects is possible by zeroing $\sin2\vartheta$. Such suppression is no longer possible if the medium birefringence varies across the beam cross section. 

   \begin{figure} [bht]
  \centering 
  \includegraphics[width=8.0 cm]{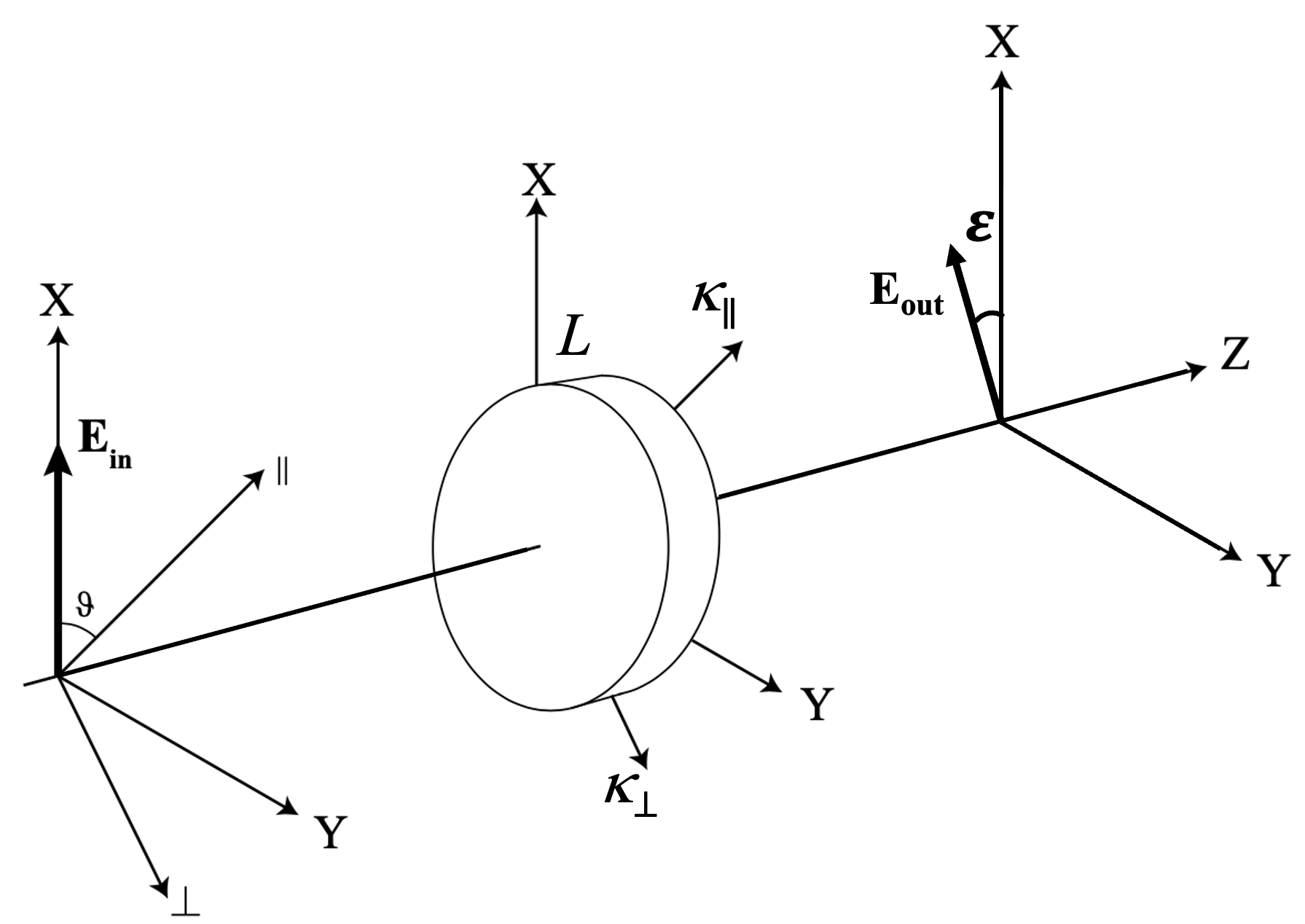}
  \caption{Reference frame for the dichroism calculations. X and Y are the axes of the laboratory frame, whereas $\parallel$ and $\perp$ are the axes of the dichroic medium, with $\kappa_\parallel<k_\perp$.  }
  \label{fig:schemadichr}
\end{figure}

Let us now come to the case of dichroism. We refer to the scheme of Figure~\ref{fig:schemadichr}. The electric field amplitude of a plane wave propagating along the Z axis in a uniform medium is described by $E(z) = E_0e^{i\frac{2\pi}{\lambda}(n+i\kappa)z}$ resulting in an exponential decay of the electric field characterised by the exponent $Q = -\frac{2\pi}{\lambda}\kappa z$. In analogy to Equation~(\ref{eq:phasedifference}), by defining a differential attenuation of the electric field $\Delta Q$
\begin{equation}
\Delta Q = -\frac{2\pi}{\lambda}\int\Delta\kappa\,dz,
\end{equation}
after crossing a length $L$ of the dichroic medium, the electric field of a linearly polarised beam of light forming an angle $\vartheta$ with the $\parallel$ direction of the dichroic medium can be approximated with
\begin{equation}
\vec{E}_{\rm out}\approx E_0\,e^{-2\pi\overline{\kappa}L/\lambda}\,
\begin{pmatrix}
          1-\dfrac{\Delta Q}{2}\cos2 \vartheta \\\\
          -\dfrac{\Delta Q}{2}\sin 2 \vartheta
      \end{pmatrix},\qquad\qquad\overline k=\frac{k_\parallel+k_\perp}{2}.
\end{equation}

The $X$ and $Y$ components of $\vec{E}_{\rm out}$ are still in phase resulting in a linear polarisation whose direction, though, is rotated by an angle
\begin{equation}
\varepsilon(\vartheta) \approx-\frac{\Delta Q}{2}\sin2\vartheta= -\frac{\pi}{\lambda}\int_L\Delta\kappa\,dz\,\sin2\vartheta.
\label{eq:rotdeltak}
\end{equation}

Like birefringence, a dichroism generates an electric field in a direction orthogonal to the input.

We show now that the key for distinguishing between these two properties of an optical material in a laboratory measurement is that ellipticity is purely imaginary, whereas rotation is a real quantity. In fact, if birefringence and dichroism are both present, the output transverse electric field is $E'_Y\approx E_0(\psi+\varepsilon)$, and the power associated with this component is $P'=P_0(\psi^2+\varepsilon^2)$. In order to distinguish between the two contributions, we add to the optical setup ellipticity and rotation modulators, denoted as $\eta(t)$ and $\varphi(t)$, respectively, in such a way that the output power contains terms linear in both $\psi$ and $\varepsilon$
\begin{equation}
P'=P'(t)=P_0\left[|\psi+\eta(t)|^2+|\varepsilon+\varphi(t)|^2\right]\approx P_0\left[\eta^2+2{\cal R}e(\psi\eta*)+\varphi^2+2\varepsilon\varphi\right].
\label{eq:extpower}
\end{equation}

If $\eta(t)$ and $\varphi(t)$ vary sinusoidally, each of the four terms in parentheses generates a different frequency component in the Fourier spectrum of $P'$.

\subsection{Birefringence and interference}

In this Section we show that a birefringence, if accompanied by interference, may generate a rotation in addition to an ellipticity, even in a non-dichroic material. This is a well-known phenomenon that has already been investigated in relation to high-finesse interferometers \cite{Zavattini2006,EPJC2016}. To illustrate this, we model the sample as an etalon in air, with air-material interfaces characterised by a power reflectivity $\mathcal{R}$  and a transmissivity $\mathcal{T}$, with $\mathcal{R}+\mathcal{T}=1$.
 Multiple internal reflections, absorption, and birefringence within the substrate are all encoded in the following Jones matrix: 
\begin{align}
\begin{aligned}
    \textbf{SI}=& {\cal T} e^{i \delta /2} e^{-\alpha L/2 } \sum_{k=0}^{\infty} [e^{i \delta}\mathcal{Z}\,\textbf{BF}^{2}]^{k} \cdot \textbf{BF} = \\ =&\mathcal{T}e^{i \delta /2} e^{-\alpha L/2 } \left[(1-e^{i \delta}\mathcal{Z}\,\textbf{BF}^{2})^{-1} \cdot \textbf{BF} \right]
    \label{eq:siliconjones}
    \end{aligned}
\end{align}
  where
  \begin{equation}
  \delta=\frac{4\pi nL}{\lambda},
  \label{eq:delta}
  \end{equation}
  the $\textit{round trip}$ phase acquired by the trapped light, determines the interference condition; $\alpha=4\pi\kappa/\lambda$ is the absorption coefficient; $\mathcal{Z}= f\mathcal{R}e^{-\alpha L}$, with $f$ the fraction of light that interferes, mimicking the finite extent of the beam accompanied by an imperfect alignment and focusing; $\textbf{BF}$ represents the birefringent element defined in Equation~(\ref{eq:bf}). The output electric field turns out to be, to first order: 
\begin{align}
\begin{aligned}
    \vec{E}_{\rm out} =  \textbf{SI}  \cdot \vec{E}_{\rm in} \approx E_0\dfrac{\mathcal{T} e^{i \delta /2 } e^{-\alpha L/2}}{1 - e^{i \delta}\mathcal{Z}}     
    \begin{pmatrix}
        1 \\ 
        \psi(\vartheta)\,\dfrac{1 + e^{i \delta}\mathcal{Z}}{1 - e^{i \delta}\mathcal{Z}}
    \end{pmatrix}.
    \label{eq:E_out}
    \end{aligned}
\end{align}

As the second component of the vector in the above equation is a complex number, we conclude that a birefringent etalon generates both an ellipticity $\Psi$ and a rotation $\Phi$:
\begin{align}
    \Psi= \dfrac{\psi(1-\mathcal{Z}^2)}{1-2\mathcal{Z}\cos\delta+\mathcal{Z}^2}\qquad\qquad\text{and}\qquad\qquad\Phi=\dfrac{i\psi\,2\mathcal{Z}\sin\delta}{1-2\mathcal{Z}\cos\delta+\mathcal{Z}^2}.
    \label{eq:ellrot}
\end{align}

\begin{figure} [bht]
  \centering 
  \includegraphics[width=6 cm]{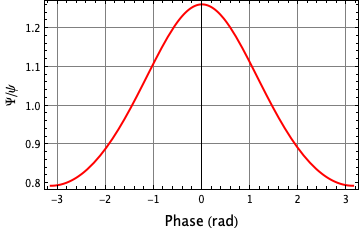}
  \hspace{0.5cm}
  \includegraphics[width=6 cm]{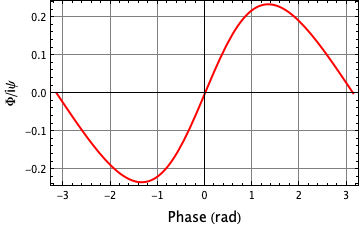}
  \caption{Ratios $\Psi/\psi$ (left) and $\Phi/i\psi$ (right) from Equation~(\ref{eq:ellrot}) as a function of phase $\delta$ for a 1-mm thick Si sample having $\mathcal{Z}=0.118$ at $\lambda=1064$~nm (see text). An experimental determination of the ellipticity curve is shown in  Figure~\ref{fig:modulazione} below.}
  \label{fig:Psi_psi}
\end{figure}

We are interested in birefringence $\Delta n$; this quantity is encoded in $\psi$, but what is measured is $\Psi$. If one knows a value for $\mathcal{Z}$, the phase $\delta$ can be then extracted from the ratio
\begin{equation}
    \frac{\Phi}{i\Psi}=\frac{2\mathcal{Z}}{1-\mathcal{Z}^2}\,\sin\delta.
    \label{eq:ratioPhiPsi}
\end{equation}
The same quantity can be extracted from the expression of the power associated with the $X$ component of the electric field (in practice  the total power transmitted by the sample)
\begin{equation}
 P_{\rm out}=P_0\,\frac{e^{-\alpha L}(1-\mathcal{R})^2}{1-2\mathcal{Z}\,\cos\delta+\mathcal{Z}^2}.
\end{equation}
Assuming $f=1$ one has
\begin{equation}
\cos\delta=\frac{1+\mathcal{Z}^2}{2\mathcal{Z}}-\frac{P_0}{P_{\rm out}}\frac{(1-\mathcal{R})^2}{2\mathcal{R}}.
\label{eq:deltafromP}
\end{equation}
More in general, an expression for $\Psi$ can then be written that avoids the need of measuring rotations:
\begin{equation}
\Psi=\psi\,\frac{P_{\rm out}}{P_0}\frac{1-{\cal Z}^2}{e^{-\alpha L}(1-{\cal R})^2}.
\end{equation}

In the literature, values of both $n_{\rm Si}$ and $k_{\rm Si}$ at RT and $\lambda=1064$~nm are found \cite{Briggs1950,Green1995,Green2008,Schinke2013,Schinke2014,Schinke2015,Franta2017}. However, while all the determinations of the first quantity converge to $n_{\rm Si}^{(1064)}\approx3.554\pm0.004$ so that
  \begin{equation}
      {\mathcal R}_{\rm Si}^{(1064)}=\left(\frac{n_{\rm Si}^{(1064)}-1}{n_{\rm Si}^{(1064)}+1}\right)^2=0.3146\pm0.0005
\end{equation}
the values of $k_{\rm Si}^{(1064)}$ are much more scattered, probably due to the strong dependence on temperature of the absorption band of silicon \cite{Green2008,Weakliem1979,Sin1984,Nguyen2014}: $\alpha_{\rm Si}^{(1064)}\approx(982\pm61)\text{~m}^{-1}$. With $f=1$, a 1-mm thick sample will then have ${\cal Z}^{(1064)}={\cal R}_{\rm Si}^{(1064)}e^{-\alpha_{\rm Si}^{(1064)} L}=0.118\pm0.007$. In this situation, shown in Figure~\ref{fig:Psi_psi}, $\Psi$ can differ from $\psi$ by at most $\approx\pm20$\%. According to Equation~(\ref{eq:delta}), a complete interference cycle corresponds to a thickness change $\Delta L\approx150$~nm, a path difference too short to alter absorption, since $\alpha_{\rm Si}\,\Delta L\approx1.5\times10^{-4}$. Employing the usual formalism of multiple interference \cite{BornWolf1980} it is also interesting to calculate the finesse ${\cal F}$ of this 1~mm Si etalon for $\lambda=1064$~nm radiation:
 \begin{equation}
     {\cal F}^{(1064)}=\frac{\pi\sqrt{{\cal Z}^{(1064)}}}{1-{\cal Z}^{(1064)}}\approx1.2
 \end{equation}
 with reflectivity and absorption equally contributing to limit this value.

 In the case of $\lambda=1550$~nm \cite{Briggs1950,Franta2017,Salzberg1957,Li1993} $n_{\rm Si}^{(1550)}\approx3.479\pm0.004$ so that
\begin{equation}
 {\cal R}_{\rm Si}^{(1550)}=\left(\frac{n_{\rm Si}^{(1550)}-1}{n_{\rm Si}^{(1550)}+1}\right)^2=0.3064\pm0.0004.
 \label{zeta1550}
\end{equation}

Moreover, one can safely assume $\alpha_{\rm Si}^{(1550)}=0$, hence ${\cal Z}^{(1550)}={\cal R}_{\rm Si}^{(1550)}$ if $f=1$, and ${\cal F}^{(1550)}=2.5$. Note that at this wavelength the difference between $\Psi$ and $\psi$ can amount to $\pm50\%$.

\section{Experimental method}\label{sec:Experimental}

\begin{figure}[bht]
  \centering 
  \includegraphics[width=13 cm]{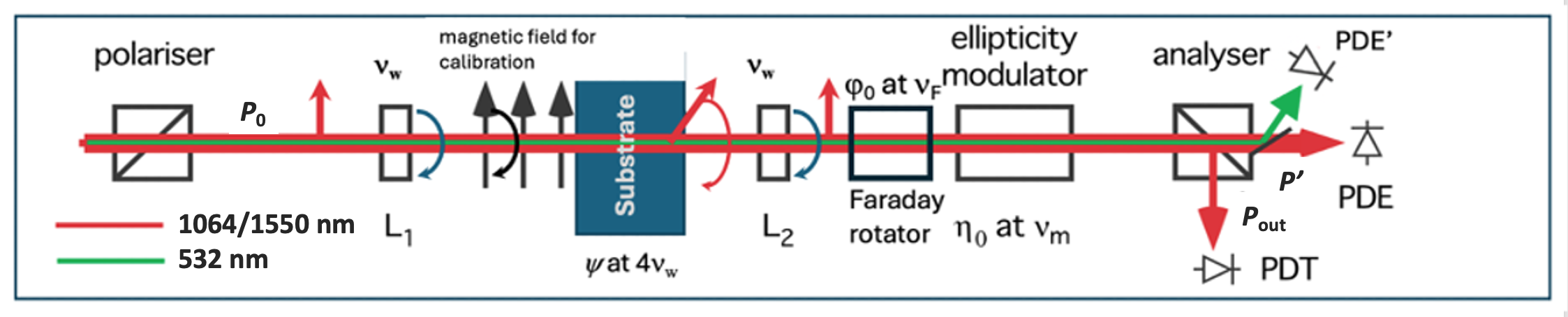}
  \caption{Polarimeter for measuring static birefringence of substrates. The two co-rotating half-wave plates (HWPs) L$_1$ and L$_2$ make the polarization rotate at a frequency $2\nu_{\rm w}$ in the space between them. A rotating permanent magnet is used for calibration purposes. A $532$~nm laser allows independent alignment of the HWPs. The setup includes rotation and ellipticity modulators (Faraday and photo-elastic modulator, respectively). PDT: photodiode for measuring the power of the $X$ component of the polarisation (total power $P_{\rm out}$); PDE: photodiode for infrared light, $Y$ component ($P'$); PDE': photodiode for green light, $Y$ component.}
  \label{fig:schemapolarimetro}
\end{figure}

The experimental setup is shown in Figure~\ref{fig:schemapolarimetro}. Birefringence measurements are performed using a frequency-tunable laser source at $\lambda=1064$~nm or, more recently, at $\lambda=1550$~nm with a newly deployed system; the 532~nm laser is used for independent alignment of each of the two HWPs. The optical path runs horizontally. The infrared light power $P_0$ ranges from 10 to 30~mW. The polarimeter extends between the polariser defining the input polarisation, and an analyser separating the $X$ and $Y$ components of the output electric field. The apparatus is hosted in a room stabilised in temperature nominally around $295\pm1$~K.

\begin{figure}[bht]
  \centering 
  \includegraphics[height=5.5 cm]{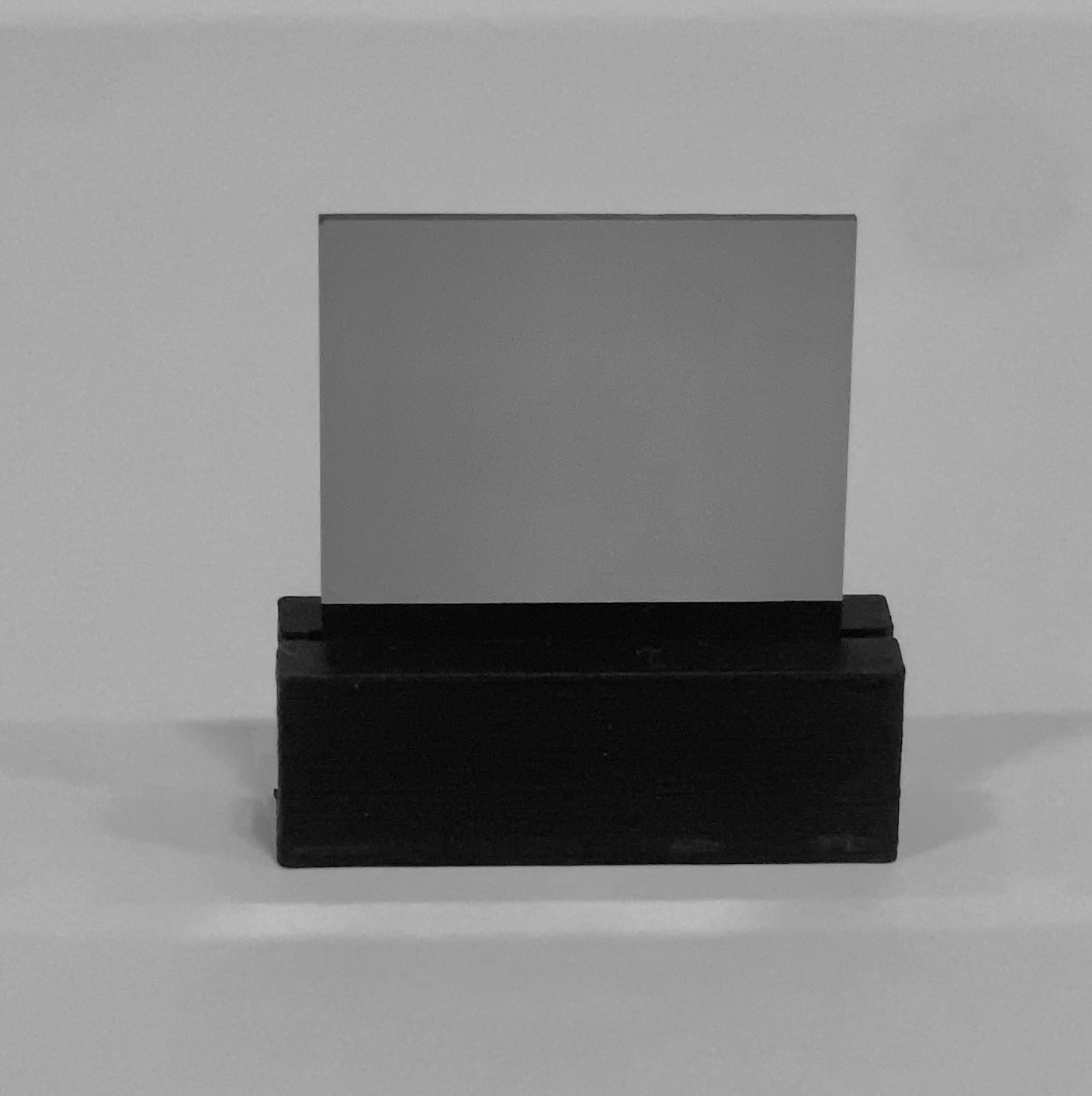}
  \hspace{0.5cm}
   \includegraphics[height=5.5 cm]{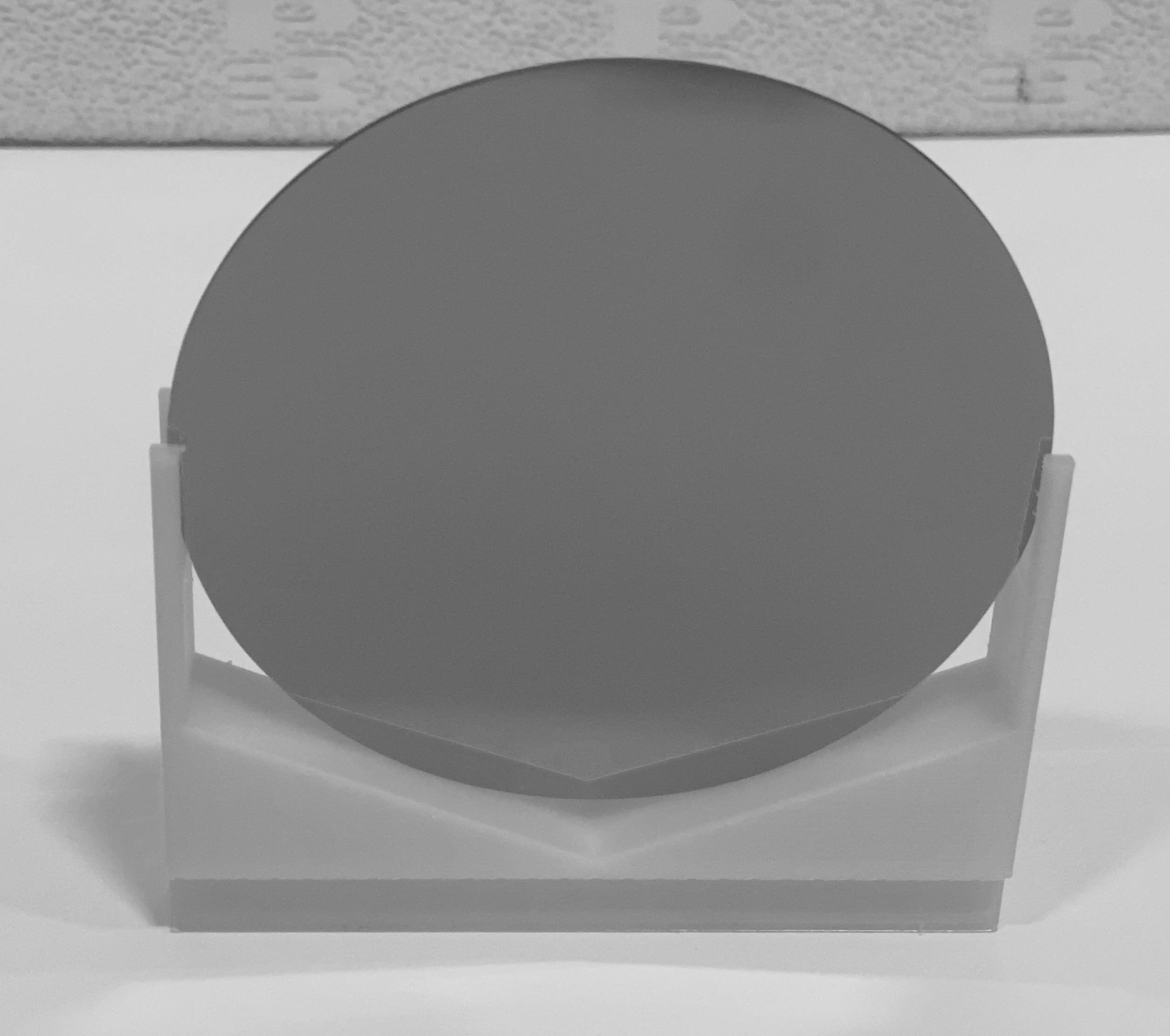}
  \caption{Pictures of the 1-mm thick samples sitting in their holders. Left: $25\times25{\rm~mm}^2$ sample. Right: 9.6~mm diameter sample.}
  \label{fig:Holders}
\end{figure}

The tested samples are 1-mm thick (100) or 2-mm thick (110) crystalline silicon wafers positioned orthogonally to the incident beam at the position of the beam waist $w_0$ ($Z=0$ coordinate of the Gaussian beam). Given this and the thinness of the samples, the approximation $f=1$ is fully justified in Equations~(\ref{eq:ratioPhiPsi}) and (\ref{eq:deltafromP}) for the thinner samples. The samples are surrounded by thermally insulating walls intended to mitigate thermal drifts due to the cycles of the air conditioning system. The samples are vertically supported, without stress by 3D printed holders as shown in Figure \ref{fig:Holders}. The bottom edge of the square samples rests on a flat surface, whereas the round wafers are supported in a v-shaped profile, thus touching the holder at two points. The holders are kept in a $XY\theta_X\theta_Y$ positioning system capable of spanning an area $25\times25$~mm$^2$ to produce a birefringence map.

\subsection{Modulation polarimetry}

\begin{table}[bht]
\centering
\caption{Fourier components of light power $P'$ of the extinguished beam: ellipticities $\Psi$ appear at frequencies $\nu_{\rm m}\pm4 \nu_{\rm w}$; rotations $\Phi$ appear at frequencies $\nu_{\rm F}\pm4\nu_{\rm w}$.}
\begin{tabular}{cccc}
\hline
Component           & Frequency                    & Amplitude$/P_{\rm out}$                     \\\hline
$\tilde{P}'_{DC}$           & DC                           & $\sigma^2+\eta_0^2/2+\varphi_0^2/2$ \\
$\tilde{P}'_{\Phi\pm}$      & $\nu_{\rm F}\pm4\nu_{\rm w}$ & $\Phi_0\varphi_0$                   \\
$\tilde{P}'_{2\nu_{\rm F}}$ & $2\nu_{\rm F}$               & $\varphi_0^2/2$                     \\
$\tilde{P}'_{\Psi\pm}$      & $\nu_{\rm m}\pm4\nu_{\rm w}$ & $\Psi_0\eta_0$                      \\ 
$\tilde{P}'_{2\nu_{\rm m}}$ & $2\nu_{\rm m}$               & $\eta_{0}^2/2$                      \\
\hline
   \end{tabular}
    \label{tab2}
  \end{table}

The ellipticity $\Psi$ induced by the substrate under test is modulated by means of two half-wave plates, L$_{1}$ and L$_{2}$, co-rotating at a frequency $\nu_{\rm w}$, resulting in the angle $\vartheta$ of Figure~\ref{fig:schemabirif} rotating in space at a frequency $2\nu_{\rm w}$. As a consequence, the ellipticity of Equation~(\ref{eq:elliptdeltan}) oscillates at a frequency $4\nu_{\rm w}$:
\begin{equation}
    \Psi=i\Psi_0\sin4\omega_{\rm w}t.
\end{equation}
This serves to move interesting signals away from DC, where the noise is generally higher. A typical value for $\nu_{\rm w}$ is $\nu_{\rm w}=2$~Hz. The same modulation is experienced by the rotation, both the ``spurious'' one of Equation~(\ref{eq:ellrot}) ($\Phi$) and the one due to dichroism of Equation~(\ref{eq:rotdeltak}) ($\varepsilon$). In what follows, though, we assume that no dichroism is present in the substrates, namely $\varepsilon=0$. Simultaneous measurements of ellipticity and rotation are performed using heterodyne detection, with a Faraday rotator that adds a rotation $\varphi(t)= \varphi_0\cos\omega_{\text{F}}t$ at a frequency $\nu_{\text{F}}$ of a few kilohertz, and a photoelastic modulator (PEM) inducing an ellipticity $\eta(t)=i\eta_0\cos\omega_{\text{m}}t$ at a frequency $\nu_{\rm m}\approx50$~kHz. As already mentioned, all the above listed time dependences generate a well-defined Fourier spectrum of the extinguished power $P'(t)$ that we now write as
\begin{eqnarray}
\nonumber P'(t)&=&P_{\rm out}\left[\sigma^2+|\Psi(t)+\eta(t)|^2+|\Phi(t)+\varphi(t)|^2\right]=\\
&\approx&P_0\left[\sigma^2+\eta^2+2{\cal R}e(\Psi\eta*)+\varphi^2+2\Phi\varphi\right].
\label{eq:extpowerinterf}
\end{eqnarray}
where $\sigma^2\lesssim10^{-7}$ is the {\it extinction ratio} of the polarisers. The main harmonics are summarised in Table \ref{tab2}. The two quantities $\Psi_0$ and $\Phi_0$ are then calculated as \cite{25yeffort}
\begin{equation}
 \Psi_0=\frac{\tilde{P}'_{\Psi-}+\tilde{P}'_{\Psi+}}{\tilde{P}'_{2\omega_{\rm m}}}\frac{\eta_0}{4}
 \qquad\qquad {\rm and}\qquad\qquad
 \Phi_0=\frac{\tilde{P}'_{\Phi-}+\tilde{P}'_{\Phi+}}{\tilde{P}'_{2\omega_{\rm F}}}\frac{\varphi_0}{4}.
 \end{equation}
These values are inserted into Equation~(\ref{eq:ratioPhiPsi}) from which $\delta$ is obtained. This number is used in turn in Equation~(\ref{eq:ellrot}) to calculate $\psi_0$ and hence $\Delta n$ from Equation~(\ref{eq:elliptdeltan}). We point out that the final result of the birefringence measurement is an amplitude and a phase allowing to determine the spatial direction of the optic axis.

In practice, the extinguished power $P'$ is collected by a high-gain, low-noise detector, and two lock-in amplifiers are employed to demodulate the signal at $\nu_{\rm m}$ and $\nu_{\rm F}$. Each amplifier folds the $\tilde{P}'(\omega)$ spectrum around its frequency, and generates in the output a single component at frequency $\nu_\Psi=\nu_\Phi=4\nu_{\rm w}$ with amplitudes $\tilde{P}'_\Psi=\tilde{P}'_{\Psi-}+\tilde{P}'_{\Psi+}$ and $\tilde{P}'_\Phi=\tilde{P}'_{\Phi-}+\tilde{P}'_{\Phi+}$, namely the sum of the amplitudes of $\tilde{P}'_{\Psi\pm}$ for the ellipticity, and of $\tilde{P}'_{\Phi\pm}$ for the rotation. 

\subsection{Systematics}

\begin{figure}[bht]
    \centering
\includegraphics[width=0.75\linewidth]{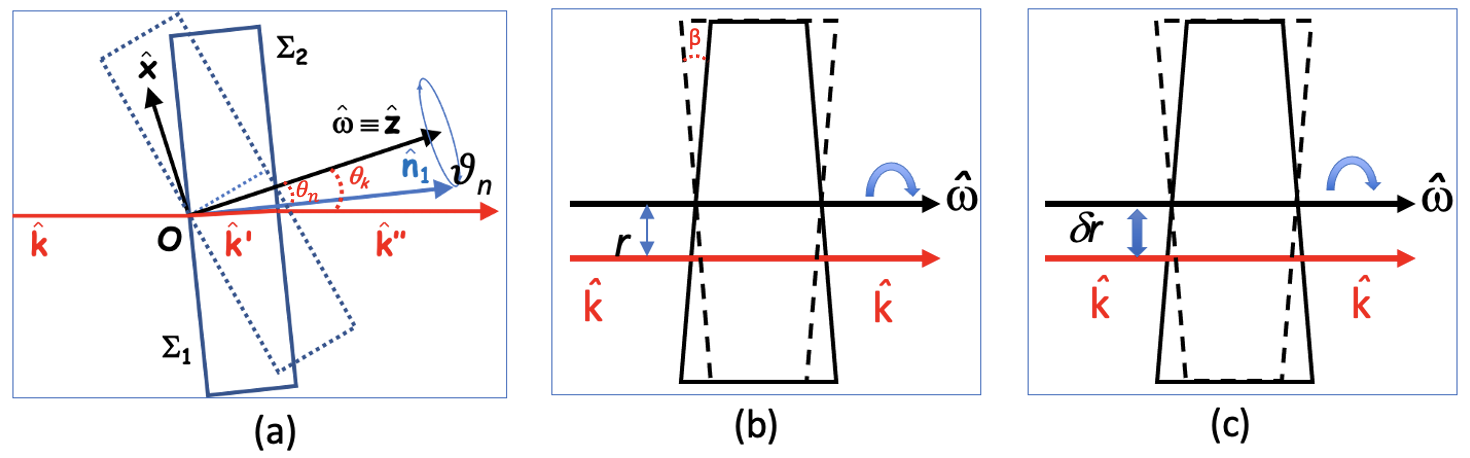}
    \caption{Different alignment issues which will generate spurious ellipticity harmonics of $\nu_{\rm w}$. In particular the leftmost and rightmost effects will generate a 4-th harmonic in $\psi(t)$ (see reference \cite{Zavattini2022}).}
    \label{fig:spuriouseffects}
\end{figure}

As discussed in reference \cite{Zavattini2022}, rotating half-wave plates generate spurious ellipticity due to a combination of optical and mechanical imperfections. The total observed ellipticity is the sum of contributions of the sample, $\Psi(t)$, and the spurious, $\psi_{\text{spurious}}(t)$, with the spurious term being in turn the sum of the contributions of the two wave plates \cite{Zavattini2022,Zavattini2016}: 
\begin{equation}
   \psi_{\text{spurious}}(t)= i\frac{\alpha_1(t)}{2}\sin\left[2\omega_{\rm w}t+2\vartheta_1\right]+i\frac{\alpha_2(t)}{2}\sin\left[2\omega_{\rm w}t+2\vartheta_2\right].
\end{equation}
Here the phase shifts $\alpha_{1}(t)$ and $\alpha_{2}(t)$ describe the instantaneous deviations of the wave plates from the ideal $\pi$-shift. The $\alpha$'s are time dependent due to the structural and alignment defects that are summarised below and thoroughly discussed in reference \cite{Zavattini2022}; their time dependence features harmonics of the rotation frequency $\nu_{\rm w}$. By expanding $\alpha_{1,2}(t)$ in terms of $\cos \vartheta(t)$, we obtain: 
\begin{equation}
\alpha_{\rm 1,2}(\vartheta,T) = \alpha^{(0)}_{1,2}(T) + \alpha^{(1)}_{\rm 1,2}\cos\vartheta(t) + \alpha^{(2)}_{\rm 1,2}\cos2\vartheta(t) + ...,
\end{equation}
where $\alpha^{(0)}_{1,2}(T)$ are the temperature-dependent time averaged deviations from $\pi$ of the retardation of the two wave plates, and $\alpha^{(1)}_{1,2}$ and $\alpha^{(2)}_{1,2}$ are deviations from $\pi$ of the retardation with a one-revolution periodicity and a half-revolution periodicity, respectively. They can be due to a wedge $\beta\lesssim10^{-6}$~rad between the front and back surfaces of each wave plate coupled to a non-centered beam; to a misalignment of laser beam, rotation axis and normal to the wave plate surface; and finally to a periodic de-centering of the wave plate during rotation. These effects are sketched in Figure~\ref{fig:spuriouseffects}.

To minimise the spurious effects, each rotating wave plate must be aligned independently but this cannot be done at the nominal wavelength of the half-wave plates: a single rotating half-wave plate does not allow a measurement in extinction. The frequency doubled 532-nm beam is used for this reason. At 532 nm, 1064-nm half-wave plates behave as full-wave plates -- with an extra contribution to $\alpha^{(0)}_{1,2}(\rm T)$ due to dispersion; in this case, then, the presence of a single rotating 1064-nm half-wave plate leaves the polarisation direction fixed, allowing to work in extinction. Special half-wave plates have been provided for the task: at $\lambda=1064$~nm the best agreement is found for a quartz thickness of 1.16~mm, corresponding to $19\lambda/2$ and $20\lambda$ for $\lambda=1064$ and 532~nm, respectively. At $\lambda=1550$~nm the coincidence is somewhat poorer but still effective, with a a 0.64~mm quartz thickness corresponding to $7\lambda/2$ retardation and $11\lambda$ at $\lambda=1550$~nm and 532~nm, respectively. In this way, at 532~nm the plates do not rotate the polarisation; the geometry-related spurious effects of Figure~\ref{fig:spuriouseffects}, though, remain. Therefore, each rotating wave plate can be independently aligned with the 532-nm beam, while the other remains still. For alignment, the rotating wave plates are mounted on an optical mount with the four degrees of freedom $XY\theta_X\theta_Y$.

A typical acceptable value for the spurious 4-th harmonic ellipticity induced by the rotating half-wave plates is $|\psi^{\rm (4-th)}_{\rm spurious}| \lesssim 10^{-5}$. This spurious ellipticity can be determined (as amplitude and phase) by removing the substrate sample from the optical path; its value has to be vector-subtracted from the value measured with the sample in place. This subtraction determines the ultimate sensitivity of the polarimeter which is estimated to be about $|\psi^{\rm (4-th)}_{\rm spurious}(t)|/5 \approx 2\times10^{-6}$ corresponding to an optical path difference sensitivity $S_{\Delta{\cal D}} = \int_{\rm sample}\Delta n\,dz \lesssim 10^{-12}$~m. However, the sample contribution is typically much larger than the spurious term: $\Psi_0\gg|\psi^{\rm (4-th)}_{\rm spurious}|$.

\subsection{Calibration}

For calibration, a rotating dipolar magnetic field is placed between the rotating half-wave plates. Two different magnets have been employed: the first 0.82-m long with $B=2.5$~T, the other 0.20-m long with $B=2.3$~T. The magnetic field $B$ makes air birefringent through the Cotton-Mouton effect \cite{Rizzo1997}: 
 \begin{equation}
     \Delta n_{\text{CM}}= \Delta n_{\rm u}\,P B^{2},
 \end{equation}
where $P$ is the gas pressure and $\Delta n_{\rm u}$ is a (non-adimensional) {\it unitary birefringence}. Assuming that air is mainly composed of 20\% oxygen and 80\% nitrogen one finds \cite{Rizzo1997}: 
\begin{equation}
\Delta n_u^{\rm(air)}\approx-6.4\times10^{-13}{\rm~T}^{-2}{\rm atm}^{-1}
\label{eq:CM}
\end{equation}
at $\lambda=1064$~nm. The field rotates at a frequency $\nu_B=0.5$ Hz, and generates a known ellipticity at a frequency of $4\nu_{\rm w}\pm2\nu_B$, where the sign depends on the relative rotation directions of the wave plates and the magnetic field. We note also that, in addition to providing a reference value for birefringence, the magnet allows also the calibration of the direction of the birefringence axis. The intensity and direction of this magnetically induced birefringence have indeed been used to verify and sometimes calibrate the experimental results presented in this work.

\section{Two-dimensional birefringence maps of crystalline silicon}\label{sec:Results}

\subsection{(100) samples}

We investigated two types of samples of crystalline (100) oriented silicon 1~mm thick: the first is a prime-quality microelectronics wafer manufactured by Siltronic (DE) that has been cut in-house in squares $25\times25$~mm$^2$; the second was cut and polished by SurfaceNet (Rheine, DE) in the form of a 96.5~mm diameter wafer extracted from the initial part of a float-zone ingot grown by IKZ (Berlin, DE) in a quest for the highest purity. The orientation of the surface of both samples has been verified by X-ray diffraction and found precise to less than $1^\circ$. The $\lambda=1064$~nm laser was used.

\begin{figure}[bht]
    \centering
    \includegraphics[width=8.0cm]{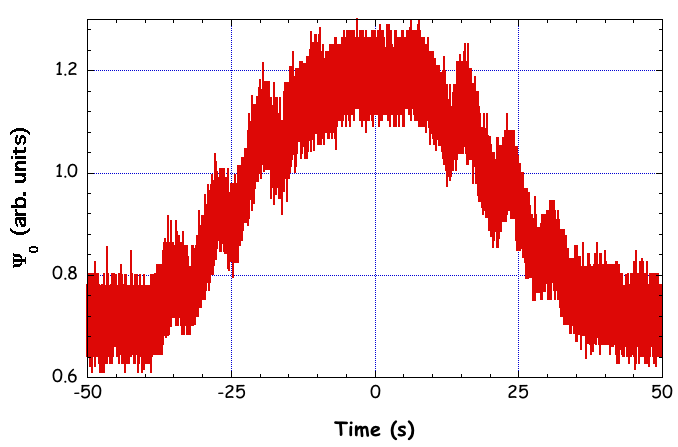}
    \caption{Elliticity $\Psi_0$ as a function of phase $\delta$ for a 1~mm mono-crystalline silicon sample; this plot is an experimental counterpart to Figure~\ref{fig:Psi_psi}. To drive the sample etalon between extremal interference conditions, a sinusoidal voltage at $\nu=8$~mHz has been applied to the laser head, thus making the emission frequency change.}
    \label{fig:modulazione}
\end{figure}

For these samples the nominal value of the $\cal{Z}$ coefficient of equation~(\ref{eq:ellrot}) has been checked by exploiting the tunability of the 1064~nm laser system. According to Equation~(\ref{eq:delta}) the $\Delta\nu=30$~GHz tunability range of the laser is about sufficient to make the ellipticity $\Psi_0$ of Equation~(\ref{eq:ellrot}) swing between a minimum $\Psi_{\rm min}=(1-{\cal Z})/(1+{\cal Z})$ and a maximum $\Psi_{\rm max}=(1+{\cal Z})/(1-{\cal Z})$, in principle thus allowing the direct measurement of $\cal{Z}$. In Figure~\ref{fig:modulazione}, the periodic modulation of the ellipticity $\Psi_0$ of Equation~(\ref{eq:ellrot}) is reported; the oscillation recorded is compatible with the value of the literature reported above, but is affected however by a larger uncertainty. As a matter of fact the value
\begin{equation}
    {\cal Z}=0.128\pm0.018
    \label{eq:zeta}
\end{equation}
deduced from the data of Figure~\ref{fig:modulazione} is the value adopted in the calculations. 

\begin{figure}[bht]
        \includegraphics[width=6cm]{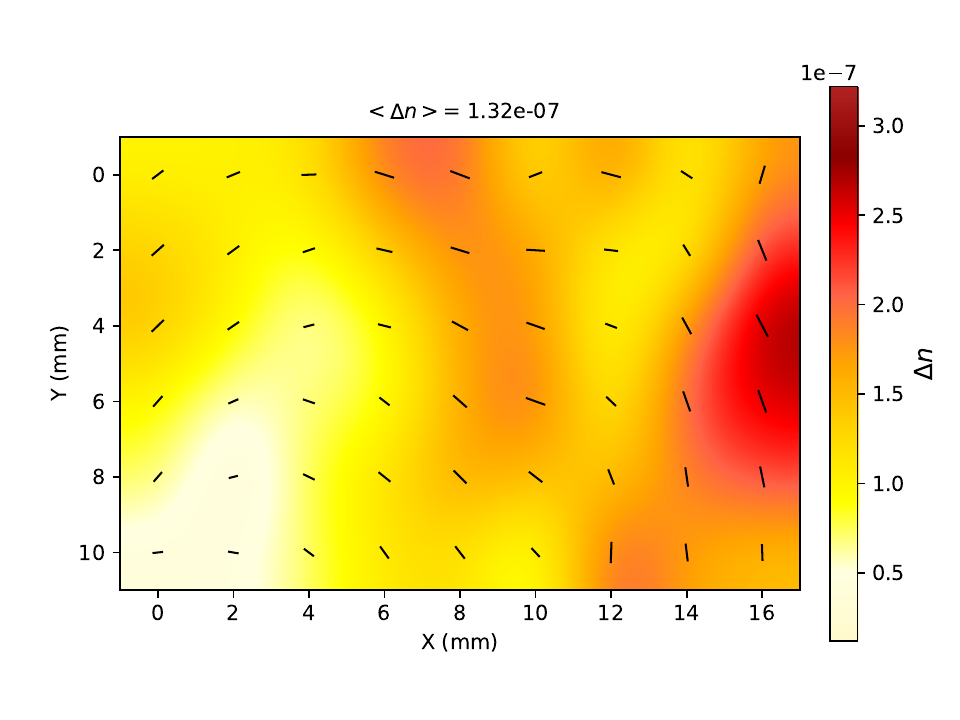} 
    \hspace{0.5cm}        \includegraphics[width=6cm]{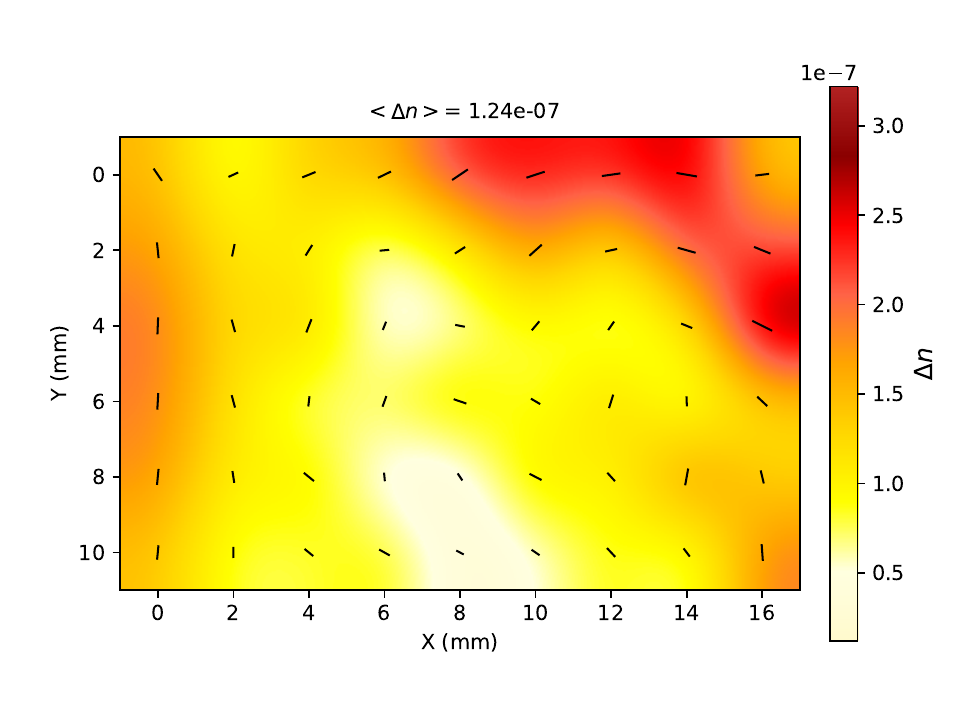}
    \caption{Birefringence maps of two 1~mm thick mono-crystalline (100) silicon samples $25\times25$~mm$^2$. Step size is $\Delta x=2$~mm: only a fraction of the samples area has been explored.} 
    \label{fig:mappa1}
\end{figure}

Birefringence maps of the in-house cut samples are presented in Figure~\ref{fig:mappa1}, with step size $\Delta x=2$~mm. In the maps, the direction of the birefringence axis is represented by a line whose length is a linear function of the magnitude of $\Delta n$; this magnitude is also encoded in the color scale. For these (100) samples the measured directions of the birefringence axis are undetermined to an additive unknown common value. The maps show a clear nonuniformity in birefringence both in value and in direction, specific for each sample. All angles are covered, indicating that there is no connection to the crystalline planes.

\begin{figure}[bht]
\includegraphics[width=6cm]{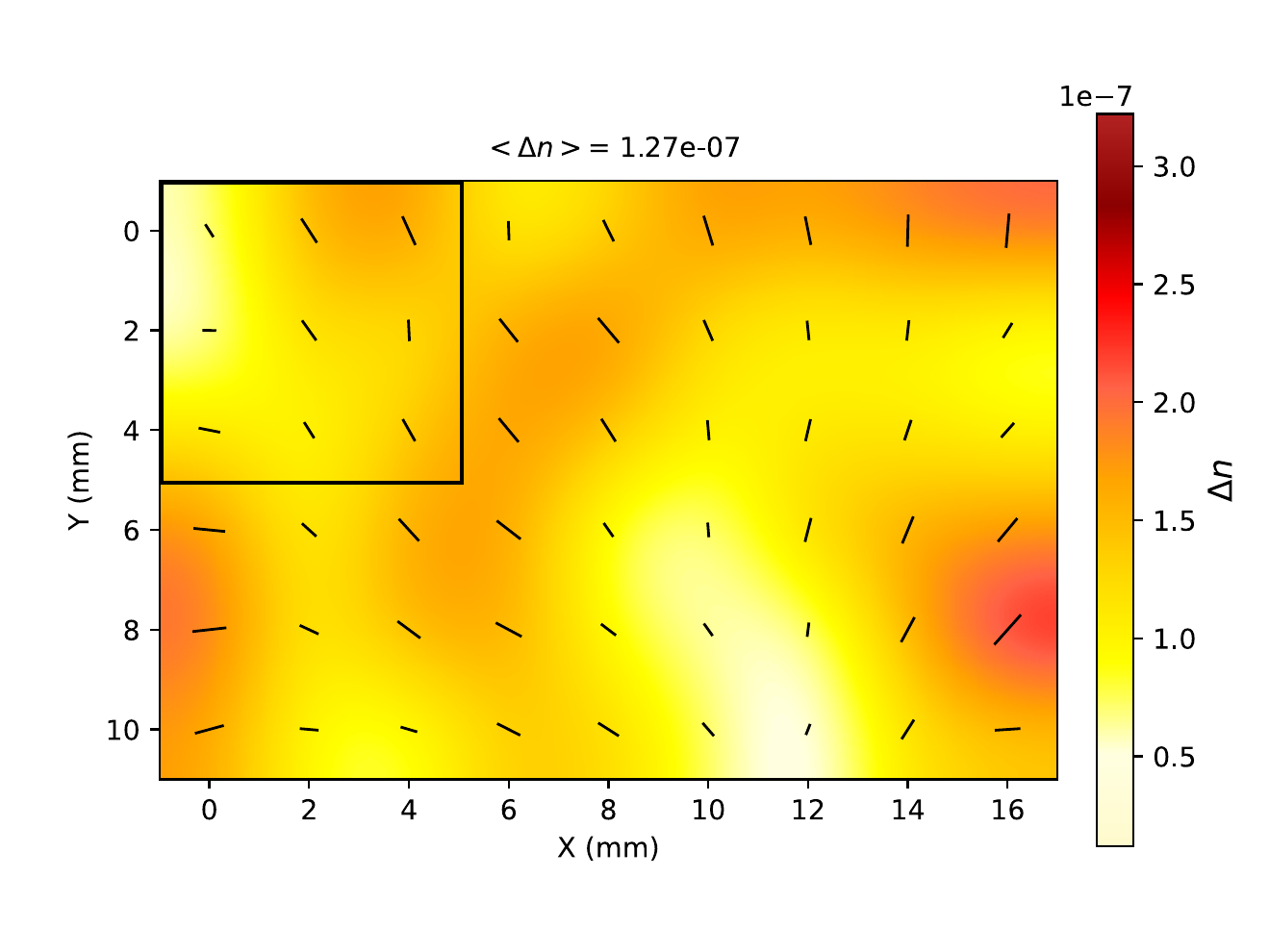} 
\hspace{0.4 cm}        \includegraphics[width=6cm]{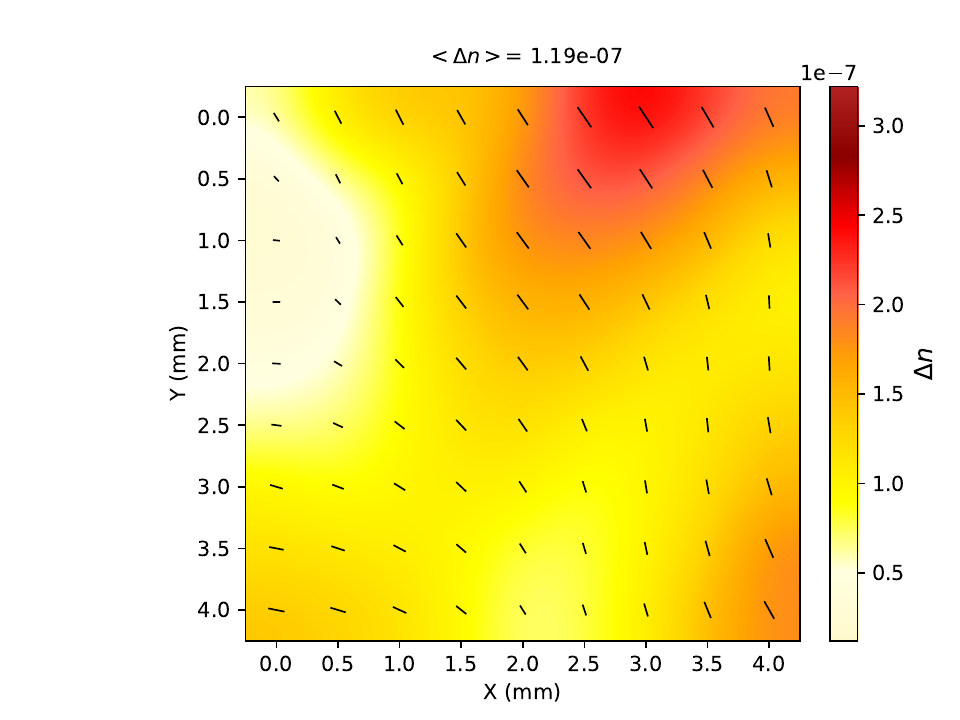}
\caption{Birefringence maps of a third 1~mm thick mono-crystalline (100) Si sample of the same batch as the two of Figure~\ref{fig:mappa1}. Left: overall birefringence map with step size $\Delta x=2$~mm (only a fraction of the sample area has been explored).  Right: zoomed-in view of the top left part of the left map, with a smaller step size of $\Delta x'=0.5$~mm. The corner measurements of the right panel replicate the corner measurements of the top left $3\times3$ square of the left panel.} 
    \label{fig:mappa2}
\end{figure}

The maps reported in Figure~\ref{fig:mappa2} are taken for a third sample of the same batch. The map on the left is analogous to the two shown in Figure~\ref{fig:mappa1}, whereas the map on the right reports the birefringence of the $4\times4$~mm$^2$ region in the top-left corner of the sample, measured with a smaller $\Delta x'=0.5$~mm step size for greater spatial resolution. The results confirm the repeatability and accuracy of the measurements, demonstrating that the observed birefringence pattern is an inherent property of the samples.

\begin{figure}[bht]
    \centering
    \includegraphics[width=8.0cm]{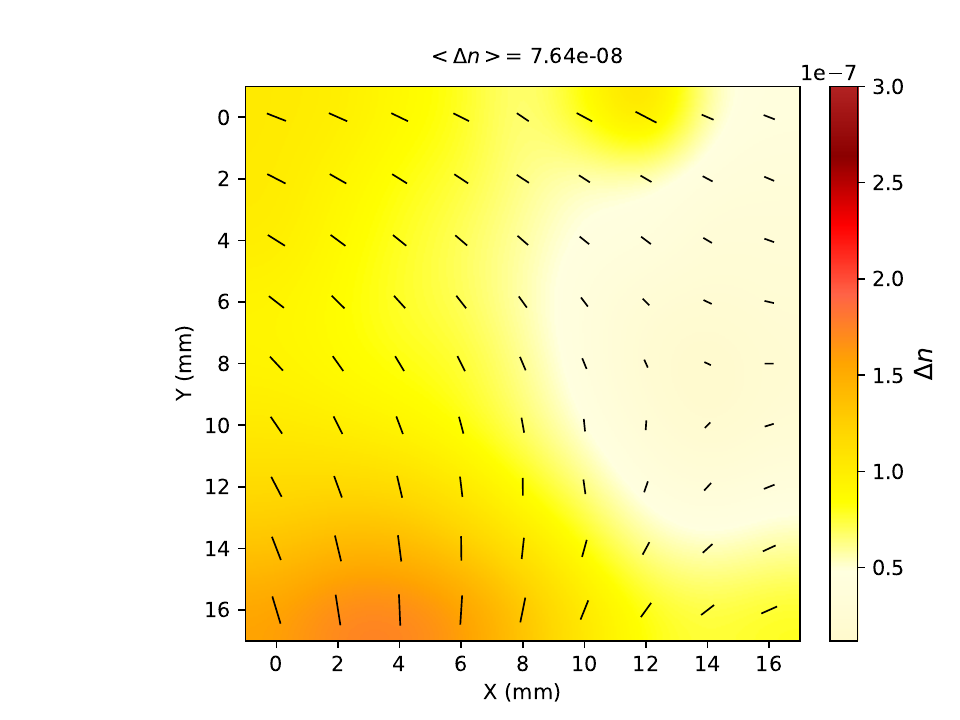}
    \caption{Birefringence map of a 1~mm thick mono-crystalline (100) Si sample with a $D=96.5$~mm diameter. Step size is $\Delta x=2$~mm: only a fraction of the sample area around the center of the disc has been explored.}
    \label{fig:sampleIKZ}
\end{figure}

The birefringence of the 1-mm thick sample grown by IKZ is shown in Figure~\ref{fig:sampleIKZ}. Unlike the in-house cut samples, in this case the birefringence is generally lower, and also the angular variation of the birefringence axis is less pronounced across the analysed area. This might depend on a better state of the surface of this sample with respect to the previous ones. We note however that, although the overall distribution appears more uniform, inhomogeneities persist across the sample.

\begin{figure}[bht]
\begin{center}
\includegraphics[width=8.0cm]{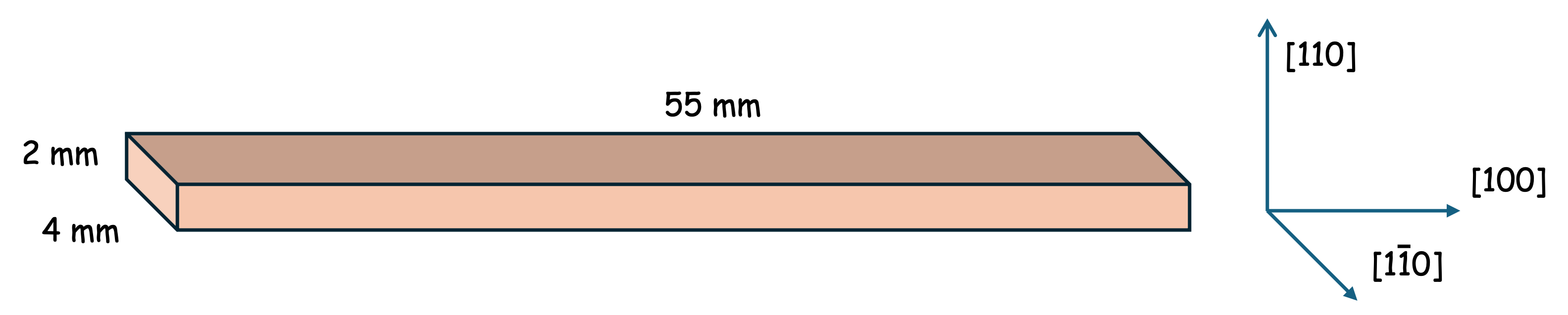} 
\caption{Geometry of the Si(110) samples.} 
    \label{fig:Silicio110}
\end{center}
\end{figure}

\subsection{(110) samples}

The (110) samples have been manufactured for channeling \cite{Kyryllin2025} by Siltronix (FR) in the form of a strip $2\times4.1\times55{\rm~mm}^3$ delimited by the planes $(110)$, $(001)$ and $(1\overline{1}0)$ (see Figure~\ref{fig:Silicio110}). These measurements were performed at $\lambda=1550$~nm. In this case we measured ${\cal Z}=0.257$ corresponding to $f={\cal Z}/{\cal R}_{\rm Si}^{(1550)}=0.84$. Birefringence has been measured at different places along the $4\times55{\rm~mm}^2$ (110) surface of the strips. The average result of the measurements is
\begin{equation}
\Delta n^{(110)}=(-1.50\pm0.15)\times10^{-6}\qquad\qquad@~\lambda=1550{\rm~nm},
\end{equation}
somewhat smaller than the theoretical prevision of Equation~(\ref{Lorentz}) and the previous measurements \cite{Pastrnak1971,Chu2002}. The direction of the fast axis was determined for this sample and was found to coincide with the [100] direction within the experimental uncertainty; this explains the negative $\Delta n$.

\begin{sidewaystable*}[bht]
\centering
\caption{Review of birefringence measurements of non-stressed mono-crystalline silicon. The values of $\Delta n_{\rm Si}^{(110)}$ found in literature have undetermined signs; in this work we find $\Delta n^{(110)}<0$. Legenda: $w_0$, beam radius; $\Delta x$, scanning step or lateral resolution; ND, not declared; NA, not applicable.  }
\begin{tabular}{ccccccc}
\hline
Surface                & $\lambda$ (nm)              & L (mm)                    & $w_0,\;\Delta x$ (mm)            & Method                            & $\Delta n_{\rm Si}$           & Ref. \\\hline
\multirow{2}{*}{(111)} & \multirow{2}{*}{1100--1200} & \multirow{2}{*}{5}        & \multirow{2}{*}{ND, NA}          & Malus + IR                        & \multirow{2}{*}{up to $10^{-4}$} & \multirow{2}{*}{\cite{Lederhandler1959}} \\
                       &                             &                           &                                  & image conversion                  & & \\\hline
(100)                  & 1150                        & 7                         & \multirow{4}{*}{ND, NA}          & \multirow{4}{*}{Malus}            & $<10^{-7}$                    & \multirow{4}{*}{\cite{Pastrnak1971}} \\
(110)                  & 1150                        & 22                        &                                  &                                   & $5.0\times10^{-6}$            & \\
(110)                  & 1450                        & 22                        &                                  &                                   & $\approx2.5\times10^{-6}$     & \\
(111)                  & 1150                        & 24                        &                                  &                                   & $<10^{-7}$                         & \\\hline
(100)                  & 1300                        & 0.5                       & ND, $<0.1$                       & Malus + scanning                  & up to $1.40\times10^{-5}$     & \cite{Fukuzawa2001} \\\hline
(110)                  & \multirow{2}{*}{1520}       & 75                        & \multirow{2}{*}{ND, $<0.1$}      & \multirow{2}{*}{Malus + scanning} & $3.20\times10^{-6}$           & \multirow{2}{*}{\cite{Chu2002}} \\ 
(100)                  &                             & 35                        &                                  &                                   & $<3\times10^{-8}$             & \\\hline 
\multirow{2}{*}{(111)} & \multirow{2}{*}{1550}       & \multirow{2}{*}{28~--~99} & \multirow{2}{*}{0.4~--~0.5, NA}  & frequency measurements            & \multirow{2}{*}{up to $1.11\times10^{-7}$} & \multirow{2}{*}{\cite{Kruger2016}} \\
                       &                             &                           &                                  & in monolithic cavities            & & \\\hline
\multirow{2}{*}{(100)} & \multirow{2}{*}{2000}       & \multirow{2}{*}{30}       & \multirow{2}{*}{ND, 1}           & linearised homodyne +             & \multirow{2}{*}{up to $5.0\times10^{-8}$}    & \multirow{2}{*}{\cite{Hamedan2023}} \\
                       &                             &                           &                                  & scanning                          & & \\\hline
(100)                  & 1064                        & 1                         & 0.5, 0.5                         & linearised heterodyne +           & up to $3.00\times10^{-7}$     & \multirow{2}{*}{This work} \\
(110)                  & 1550                        & 2                         & 0.2, NA                          & scanning                          & $-1.50\times10^{-6}$          & \\\hline
\end{tabular}
\label{Table}
\end{sidewaystable*}

\subsection{Discussion}

Table~\ref{Table} reports all the measurements of the birefringence of silicon single crystals existing in the literature, in chronological order, the first measurement dating back to 1959 \cite{Lederhandler1959}. The results for orientations (100) and (111) describe a downward trend, possibly due to an increase in material quality and increasing attention to stress-free mounting. Our results show instead somewhat larger values that cannot be attributed to systematic effects of the polarimetric method, which are about a hundred times smaller. There are a few possible reasons for this. The first is that each sample obviously has a different built-in stress which depends on past history. In particular none of our samples were produced for minimizing the intrinsic birefringence necessary for Einstein Telescope. In addition, we note that the samples studied here are very thin, which might indicate that surfaces are more birefringent than bulk.

We note also that for the measurements presented here we expect that gravity plays a role. In fact, disregarding crystal directions, silicon has a stress-optic coefficient of order $C_{\rm so}\approx2\times10^{-11}{\rm~Pa}^{-1}$ \cite{He2004}. With a mass density $\varrho_{\rm Si}=2.33\times10^3{\rm~kg/m}^3$, at the center of the $D=96.5$~mm diameter wafer the gravity-induced birefringence is
\begin{equation}
    \Delta n\approx\frac{\pi D\,\varrho_{\rm Si}\,g}{4}C_{\rm so}=3.5\times10^{-8}.
\end{equation}
No attempt has however been made to correct the results for this effect. Note that gravity has no effect on test masses suspended at the midline \cite{Kruger2016,Hamedan2023}.

As far as the (110) birefringent surface is concerned, we note that silicon has indeed multiple built-in birefringence axes that do not have an effect on the (100) or (111) planes due to perfect compensation of the six $\{110\}$ planes. We note however that varying the alignment of the laser beam by $\Delta\theta=45^\circ$ from a [100] to a [110] direction, makes the birefringence grow from zero to the value $|\Delta n_{(110)}|=1.5\times10^{-6}$. This value, albeit small, makes alignment a fundamental issue. Keeping birefringence smaller than $10^{-8}$ requires aligning the [100] crystal direction to better than 3~mrad with respect to the beam inside the crystal.

\section{Conclusions and future prospects}

In this paper we presented some measurements of the intrinsic birefringence $\Delta n$ of Si single crystals obtained with a dedicated high-sensitivity polarimeter capable of appreciating an optical path difference $S_{\Delta{\cal D}}\approx10^{-12}$~m. The results on thin silicon wafers show average birefringence values $\langle\Delta n\rangle\approx10^{-7}$, a value seemingly too high for the LF next generation gravitational wave interferometers. None of the samples, though, were produced for minimising the intrinsic birefringence. The measurements are presented as birefringence maps of small portions of the surfaces. The birefringence of Si (110) has been also measured at $\lambda=1550$~nm.

In the context of the Einstein Telescope, we also address the importance of carefully studying the impact of both the magnitude and uniformity of birefringence. If the birefringence axis is uniform in direction across the optical element, the effects can be minimised by aligning the input polarization with this axis. However, if the birefringence direction is non-uniform and its magnitude also varies spatially, the electric field will experience phase shifts that differ from point to point, making it impossible to suppress these effects. Therefore, in addition to specifying the acceptable value of birefringence, tolerances for its spatial variation must be defined, both in terms of changes in magnitude and orientation, together with the distance over which these variations have a significant impact on performance.

Our next step is to upgrade the working scheme by automating the birefringence measurement process. Furthermore, a new translation stage will be integrated allowing for larger areas mapping.

\backmatter

\bmhead{Acknowledgements}

Funded by the European Union - Next Generation EU, Mission 4 component 1 CUP F53D23001290006.

\bibliography{ETFEPaper1.bib}

\end{document}